\documentclass{iaus}
\usepackage{graphicx}

\title[Barium abundance gradient ] 
{Bimodal chemical  evolution of the Galactic disk 
and the Barium abundance of Cepheids}
\author[ L\'epine et al.]   
{Jacques R.D. L\'epine$^1$, Sergei  Andrievky$^2$, Douglas  A. Barros$^1$, Thiago  C. Junqueira$^1$ \&
  Sergio  Scarano Jr.$^3$.}

\affiliation{$^1$Instituto de Astronomia, Geof\'isica e Ci\^encias Atmosf\'ericas,
 Universidade de S\~ao Paulo, Cidade Universit\'aria, 05508-090, SP, Brasil,
  email: {\tt jacques@astro.iag.usp.br}, 
$^2$Department of Astronomy and Astronomical Observatory, Odessa
   National University, Shevchenko Park, 65014 Odessa, Ukraine,
$^3$Universidade Federal de Sergipe - Departamento de F\'isica DFI/CCET. Rod. Marechal Rondon s/n, 49.100-000, Jardim Rosa Elze, S\~ao Cristov\~ao, SE, Brazil}

\pubyear{2013}
\volume{298}  
\pagerange{xxx-xxx}
\jname{Setting the scene for GAIA and LAMOST}
\editors{S. Feltzing, G. Zhao, N. Walton \& P. Whitelock, eds.}
\begin{document}

\maketitle

\begin{abstract}
In order to understand the Barium abundance distribution in the Galactic disk based on Cepheids, one must first be aware of important effects of the corotation resonance, situated a little beyond  the solar orbit. The thin disk of the Galaxy is divided in two regions that are separated by a barrier situated at that radius. Since the gas cannot get across that barrier, the chemical evolution is independent on the two sides of it. The barrier is caused by the opposite directions of flows of gas, on the two sides, in addition to a Cassini-like ring void of HI (caused itself by the flows). A step in the metallicity gradient developed at corotation, due to the difference in the average star formation rate on the two sides, and to this lack of communication between them. In connection with this, a proof that the spiral arms of our Galaxy are long-lived (a few billion years) is the existence of this step. When one studies the abundance gradients by means of stars which span a range of ages, like the Cepheids, one has to take into account that stars, contrary to the gas, have the possibility  of  crossing the corotation barrier. A few stars born on the high metallicity side are seen on the low metallicity one, and vice-versa.  In the present work we re-discuss the data on Barium abundance in Cepheids as a function of Galactic radius, taking into account the scenario described above. The [Ba/H] ratio, plotted as a function of Galactic radius, apparently presents a distribution with two branches in the external region (beyond corotation). One can re-interpret the data and attribute the upper branch to the stars that were born on the high metallicity side. The lower branch, analyzed separately, indicates that the stars born beyond corotation have a rising Barium metallicity as a function of Galactic radius.

\end{abstract}

\firstsection 
\section{Introduction}

The ``main stream" chemical evolution models do not recognize the existence of a step of 0.3 dex in the metallicity gradient of the Galactic disk near the solar orbit radius. Moreover, since in the literature one can find many observational papers that ignore this step as well, and fit the radial gradient of the disk by a straight line across the entire range of radius, everything seems to be in order. We have to agree that this step is not easily observed; errors in the measurements and the migration of stars tend to hide it a little bit.
However, this step in the gradient reveals the existence of a major effect of the corotation resonance in the disk of the Galaxy. One cannot understand the basic physic that govern the chemical evolution of the disk without taking 
this resonance into account. We shall first describe the main effects of this resonance before proceeding to the discussion of the Barium abundance distribution.

\section{The corotation resonance as a barrier between two independent worlds}

The corotation resonance is the place where the rotation speed of the spiral pattern coincides with the rotation speed of the material of the disk. In our Galaxy both are well known. They  are shown in Figure 1, in linear (not angular) velocities. Since the spiral pattern rotates with constant angular velocity, it appears in the figure as a straight line with positive slope. The corotation radius lies only slightly beyond the solar radius.

\begin{figure}
 \centering
 \includegraphics[scale=0.48]{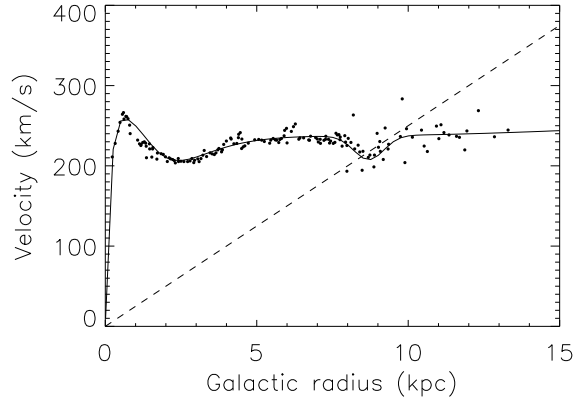}  
 \caption{The rotation curve of the Galaxy. The data points are CO observations by Clemens (1985) fitted by a smooth function. The rotation speed of the spiral pattern is indicated by a dashed line. The adopted galactic parameters here are $r_0$ =7.5 kpc,$V_0$= 235 km/s and $\Omega_p$ = 25 km/s/kpc. The rotation curve shows a minimum close to the corotation radius, which is itself a consequence of the presence of the resonance.}
\label{fig:fig1}
\end{figure}

\begin{figure}
 \centering
 \includegraphics[width=0.4\textwidth]{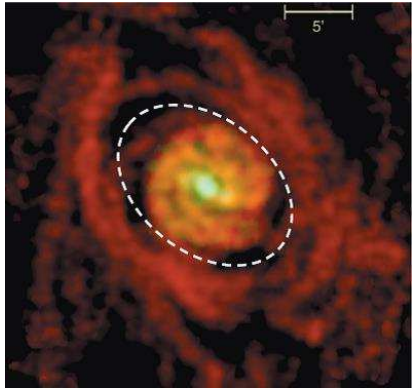}
 \caption{The HI surface density of M83 (NGC5236), according to Crosthwaite et al. (2002). At the center, the CO surface density is also shown. The dashed ellipse represent the corotation circle projected on the plane of the galaxy. One can see that it coincides with a minimum of HI density.}
\label{fig:}
\end{figure}

What happens at corotation is that the relative velocity of the gas with respect to the spiral arms reverses its direction. Consequently, the radial flux of gas induced by the spiral potential perturbation also reverses its direction. The gas of the disk flows outwards in the outer regions and inwards in the inner regions (inside corotation). Such flows are observed in external galaxies. For instance Elmegreen et al. (2009), wrote: ``In summary, the gas in NGC 1365 is observed to stream outward outside of corotation and inward inside of corotation, as expected from numerous models and observations of other galaxies".
A consequence of the gas flows in opposite directions is the formation of a ring void of gas at corotation. This is illustrated in Figure 2 in the case of M83. The figure, taken from Crosthwaite et al. (2002) shows the gas distribution in the disk of that galaxy;  we superimposed on it a dashed  ellipse which indicates  the position of the corotation radius. We see the void of gas at that radius.

Precisely at the same radius, in M83, a step has been observed in the Oxygen abundance gradient measured by 
Bresolin et al.(2009), as reproduced in Figure 3. The reason for the step is that the chemical evolution of the gas on one side of corotation is independent from that of the other, because there is no contact between them; the ring void of gas and the gas flow in opposite directions constitute a barrier. And since the star formation rate is larger inside corotation, the rate of growth of metallicity is larger in the inner region. After a few billion years, a difference in metallicity of a few dex builds up. For a discussion of the radius of corotation of M83 see  Scarano \& L\'epine (2013).

The case of M83 is not an isolated one. Scarano \& L\'epine (2013)  collected the corotation radii of a sample 
of galaxies from the literature, and looked for the presence of breaks or steps in the gradients of Oxygen abundances.
A strong correlation was found between the corotation radii and the radii of the breaks. The presence of a break
(or change of slope of the gradient) is also an indication of independent evolution on the two sides of corotation.

\begin{figure}
 \centering
 \includegraphics[scale=0.4]{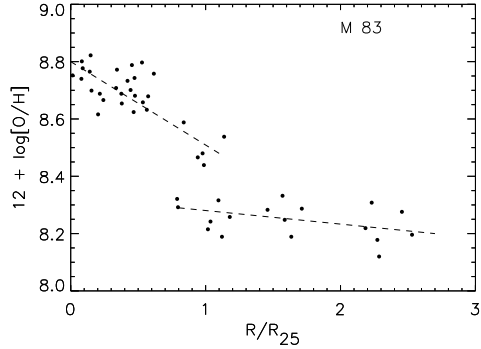}
  \caption{The Oxygen abundance as a function of radius in M83,  measured by Bresolin et al. (2009).
   The galatocentric distances are in units of R25, equal to 8.4 kpc according to the authors.
    The O abundance was measured in  HII regions of the galaxy.}
\label{fig:fig3}
\end{figure}

\vspace{-5 mm}
\section{The corotation resonance in our Galaxy}
The corotation radius of our Galaxy has been determined by many different methods, with good agreement between them; see for instance a list in the paper by Junqueira et al (2013). Of course, one can find in the literature a few discrepant values, but these are based on indirect methods (like N-body simulations, for instance) that depend on many hypotheses and uncertain input parameters. Only direct methods must be considered here. One example of such a method is the integration of the orbits of the open clusters backwards to their birthplace in the spiral arms (Dias \& L\'epine, 2005) which gave the result $R_c$ = 1.08 $\pm$ 0.06 $R_0$.

\begin{figure}
 \centering
 \includegraphics[scale=0.4]{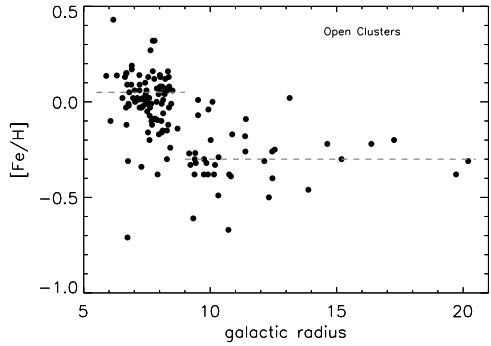}
  \caption{The Fe abundance of Open Clusters as a function of Galactic radius, taken from L\'epine  et al. (2011).
  It is possible to see the gap in the density of clusters at corotation (about 8.5 kpc for $R_0$ = 7.5 kpc), and the step down in metallicity 
  at the same radius.}
\label{fig:fig4}  
\end{figure}

The existence of a ring void of gas at the corotation radius was shown by Amores et al. (2009). The technique used was to observe the presence of very deep minima in the HI spectra of the LAB survey, that reach almost zero antenna temperature. By computing the kinematic distances of such minima, these authors showed that they are distributed along a circle situated slightly beyond the solar circle. Note that the paper did not present a map of HI, but only the position of the voids for a large number of line of sights over the whole range of longitudes. This result is in principle more robust than a map. Maps are usually based on kinematic distances of the peaks that appear in HI the spectra, and are constructed ignoring that the width of the peaks in the spectra are largely due to turbulent velocities, not to the physical widths of the arms. The circular shape of the ring void of gas shows that it cannot be interpreted as an inter-arm region.

Like in M83, the ring-shaped void of gas is associated with a step in the metallicity distribution. This is shown in Figure 4, where the Fe abundance of Open Clusters is plotted as a function of Galactic radius. One can also see in this figure the gap in the density distribution of the Open Clusters, at corotation. A simple explanation for the gap is that the clusters cannot be born in a region with very low gas density. A more sophisticated cause will be discussed below. Note that the existence of the step in the metallicity distribution of the clusters was discovered by Twarog et al. (1997), but no good explanation for it was available at that time.

\begin{figure}[h]
 \centering
 \includegraphics[scale=0.5]{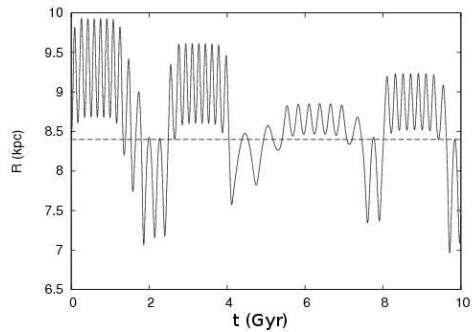}
  \caption{The radial motion of a star born close to the corotation resonance. The star alternates 
  between regions inside and outside corotation, but passes quickly at the exact resonance radius. This behaviour 
  contributes to the formation of a minimum of stellar density at corotation. }
\label{fig:fig5}
\end{figure}

 \begin{figure}[h]
 \vspace{5 mm}
  \centering
  \includegraphics[scale=0.5]{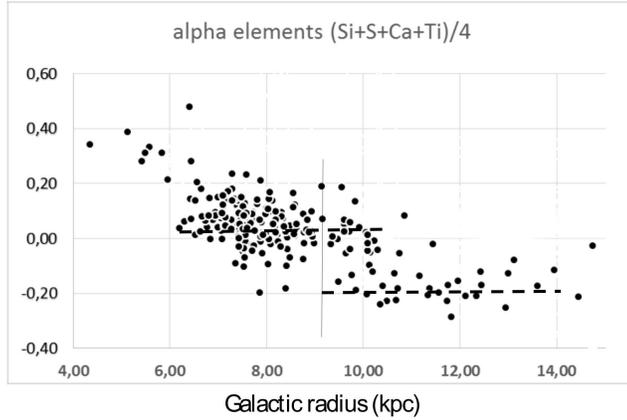}
  \vspace{-20 mm}
  \caption{The abundance of $\alpha$-elements in Cepheids as a function of Galactic radius. The average 
  of the abundances of the elements Si, S, Ca and Ti, normalized to the Solar abundances, are shown.}
  \label{fig:fig6b}
 \end{figure}

 \begin{figure}[h]
 \vspace{15 mm}
  \centering
  \includegraphics[scale=0.5]{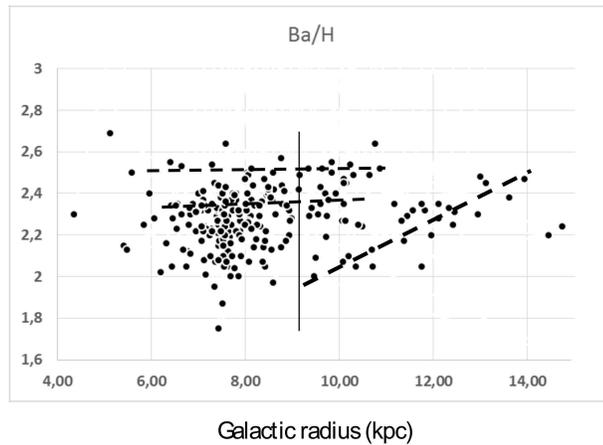}
  \vspace{-5 mm}
   \caption{The Barium abundance in Cepheids as a function of Galactic radius. The lower branch on the right side
   is the one that really represents the stars born on that side, and therefore, the abundance of the local gas. One concludes that
   the Barium abundance increases with Galactic radius, beyond corotattion. In the last two figures the corotation radius is about 9 kpc.}
 \label{fig:fig7}
 \end{figure}
 
 \vspace{-7 mm}
\section{Stars are forced to cross the corotation}

Figure 5 shows the radial motion of a star born close to the corotation radius, in this case, on the external side.
This result was obtained by integrating the orbit of the star in the presence of a spiral potential perturbation,
using the new description of this potential proposed by Junqueira et al. (2013). The corotation resonance acts on
the stars that are in its neighborhood, making them to cross the resonance from time to time, with a short crossing
time compared to the time that the star stays on each side. This behaviour  maintains a smaller
stellar density in a ring around the resonance. A detailed study of the stellar orbits near corotation  was performed
by Barros et al. (2013). 

\vspace{-5 mm}
\section{Re-visiting the Barium abundance gradient}

 The Barium abundance gradient in the Galaxy was recently investigated by Andrievsky et al. (2013) based on a
 large sample of Cepheids (270 stars). That work is a continuation of a long term effort conducted by
 Andrievsky and collaborators to investigate chemical abundances in the Galaxy (see the references 
 in that last paper). The conclusion of the paper was that the Ba abundance gradient becomes flat in the outer parts
 of the disk. We present here an alternative conclusion based on the ideas presented above.
 Figure 6 shows the gradient of  $\alpha$-elements in the Galactic disk, based on the same series of papers. We took
 an average of 4 elements in order to reduce the scattering due to errors of measurements.
 On the right  of the corotation resonance  one can see two branches in the abundance distribution.
 The upper branch with larger metallicities is due to the Cepheids that were born on the left
  side and have moved to the right side as explained in the previous section. The lower branch (smaller
  abundances) have the real abundances that correspond to the local gas where they are. One can see
  that the real gradient of the low metallicity side is flat. The overlap of the two sets (high and
  low metallicity) over a range of radius is due to the migration of stars. Note that this is totally different
   from the small overlap seen in Figure 3. HII regions have so short lifetimes that they do not migrate; in that case
  the overlap is related to a small error in the radial distances of the HII regions due to the choice of
  the inclination of M83.
  
  The Barium gradient is shown in Figure 7. By similarity with Figure 6, we recognize the set of stars 
  that were born on the left side (Galactic radii smaller than corotation) and constitute the upper branch
  on the right side. The lower branch is due to stars that really represent the local (low) abundances.
  If one focus on the lower branch, one can see that  beyond corotation the Barium abundance 
  in the gas increases 0.13 dex/kpc.
  
  This result is possibly explained by Travaglio et al. (1997).
  The r s-process distribution of elements strongly depends on stellar metallicity in the interval of [Fe/H]
  from 0 to -0.2; the more metal poor stars tend to produce slightly more baryons. Another hypothesis
  is the possible excess of AGB stars (which produce Barium) with respect to hydrogen gas, since the H density
  decreases with Galactic radius, while the AGB stars can reach large radii due to migration.



\end{document}